\documentclass[aps,pra,twocolumn,a4paper]{revtex4-1}
\usepackage{mathrsfs,amsmath, amsthm, amssymb}
\usepackage{graphicx}
\usepackage{xcolor}
\usepackage{ulem}

\begin{document}
\title{Fourth-order moment of the light field in atmosphere}

\author{R.~A.~Baskov}
\affiliation{Institute of Physics of
the National Academy of Sciences of Ukraine,\\
pr. Nauky 46, Kyiv-28, MSP 03028, Ukraine}

\author{O.~O.~Chumak}

\affiliation{Institute of Physics of
the National Academy of Sciences of Ukraine,\\
pr. Nauky 46, Kyiv-28, MSP 03028, Ukraine}

\begin{abstract}
The quasiclassical distribution function for photon density in the phase space is obtained from solution of the kinetic equation. This equation describes propagation of paraxial laser beams in the Earth atmosphere where ``collision integral" embodies the influence of turbulence. The anisotropy of photon distribution is shown and explained. For long propagation path, the explicit expression for fourth-order moment of light is obtained as a sum of linear and quadratic forms of the average  distribution function. This moment describes a spatial correlation of four light waves, giving the information about the photon distribution in the cross-section of the beam. The fourth moment can be measured using two small detectors outside the central part of the beam. The linear form describes the shot noise (quantum fluctuations) of photons. The range where the shot noise exceeds the classical noise is found analytically. Derived photon fluctuations are the valuable source of  information about statistical properties and local structure of the laser radiation that can be used for applications.

\end{abstract}

\maketitle

\section{Introduction}

Propagation of laser beam in atmosphere is crucial part for current developments of classic and quantum communication and remote sensing systems. Among others description of influence of turbulence is important for such areas as quantum key distribution \cite{veloresi, usenko}, satellite-ground communication \cite{Hosseinidehaj, Aspelmeyer}, quantum teleportation \cite{ma,ren,Hofmann}, propagation of entangled and squeezed states \cite{ursin,yin,Peuntinger,vasylev2016}, and ghost imaging \cite{wang,Shi}. In all these cases changes of spatio-temporal properties of initial radiation due to presence of atmospheric turbulence plays vital role.

Laser beam characteristics can be significantly modi\-fied  in the course of light propagation in atmosphere. Atmospheric eddies generate random fluctuations in the air refractive index (RI)  which cause an additional beam broadening as well as a  wandering effect \cite{Tatarskii1,Fante1,Fante2,Kravtsov,Gorshkov}. The results of individual beam-eddy collisions depend on their characteristic dimensions. In the case of prevailing small-size eddies, the beam broadening is observed. In contrast, for the case of small beam radius, the light can be deflected as a whole showing the effect of wandering. Despite the very small fluctuations in RI, their impact on the beam evolution can be quite significant because of accumulating effect which occurs along the propagation path. Wide dispersion of eddies parameters and index-of-refraction structure constants $C^2_n $ complicate the possibility of  numerical research for these systems. 

A spatio-temporal distribution of the radiation intensity can be obtained using different approaches. The situation is much more intricate for  the description of fluctuations of light intensity. The fluctuations are described by four-wave correlation functions (fourth-order moments of optical field) of propagating radiation. In the course of propagation, the initially coherent laser radiation acquires some properties of Gaussian statistics \cite{DeWolf}  because of the effect of atmospheric turbulence. This  complicates a theoretical analysis of fluctuation characteristics.

In early studies, the equations governing evolution of fourth-order moments were proposed \cite{Fante1,Fante2}. However, usefulness of these equations is questionable because of their complexity. Later on, the alternative approach, based on the photon distribution function (PDF) and the kinetic equation for it, was used \cite{berm,berm2009,baskov2016}. This function was defined as the photon density in the phase space. For long-distance propagation the effect of atmospheric eddies is described via random force in the kinetic equation. Obviously, this classical force cannot adequately consider the collisions which are accompanied by a substantial change of the photon momentum. In the paper  \cite{baskov2018} a more general approach which overcomes this restriction was proposed. The theory was modified, providing more general first-principle approach in order to account for the effect of photon-eddy interactions by means of collision integral. Moreover, a random nature of photon-eddy collisions is  accounted for by the Langevin source with known statistical properties \cite{Kogan,Tom}. As a result, a linear Boltzmann-Langevin equation  describes both the average and fluctuating parts of photon distribution. 

The goal of this paper is to derive asymptotical value for the fourth moment. It is highly demanded in bunch of applied researches such as intensity correlation \cite{Vellekoop,Newman}, enhanced focusing \cite{Popoff,Vellekoop2}, different imaging problems \cite{Katz, Hardy, Zhang_imag}. The fourth moment problem is simplified if the light beam is affected by multiple collisions with turbulent eddies leading to the Gaussian (normal) statistics of radiation. In this case the fourth moments can be expressed in terms of the average distribution function.

The remainder of this paper is organized as follows. In Sec. \ref{sec:pdf}, we obtain the fourth-order moment expressed in terms of average distribution function. In Sec. \ref{sec:BLE}, the kinetic equations for the average distribution function and its fluctuations are briefly analyzed.
Section \ref{sec:average_pdf} contains technical details of analytical solution for the PDF
and may be skipped by the uninterested reader. The explicit form for PDF is used for obtaining the fourth-moment asymptotics in Sec. \ref{sec:forth_moment}. The derived fourth moment is applied to estimation of aperture-averaged scintillations. The results of calculations are analyzed in Sec. \ref{sec:discussion}.  In Appendix \ref{appendix2}, the asymptotics of distribution functions, which are solutions of two different kinetic equations, are compared.

\section{Distribution function of photons and long-distance fourth-order moment for field.}
\label{sec:pdf}

The photon distribution function, defined by analogy with  distribution functions 
in physics of solids is given by \cite{UJP}

\begin{equation}\label{1threee}
\hat{f}({\bf r},{\bf q},t)=\frac 1V\sum_{\bf k}e^{-i{\bf k\cdot r}}b^\dag_{{\bf
q}+ {\bf k}/2}b_{{\bf q}-{\bf k}/2},
\end{equation}
where $b^\dag_{\bf q}$ and $b_{\bf q}$ are quantum amplitudes of  bosonic photon field  with the wave vector ${\bf q}$; $V\equiv L_xL_yL_z\equiv SL_z$ is the normalizing volume. The system should be large enough to prevent the effect of boundary conditions on beam propagation. All operators are given in the Heisenberg representation. The laser beam  propagates in the $z$ direction. Only small divergence of the beam components is assumed, i.e. the paraxial approximation holds. In this case,  the initial polarization of light remains almost unaffected  for a wide range of propagation distances (see Ref. \cite{stroh}).

The operator $\hat{f}({\bf r},{\bf q},t)$, describing the photon density in the $({\bf r},{\bf q})$ space, after summing over all values of $ \bf q $ gives the density of photons only in the spatial domain
\begin{equation}\label{2my}
\hat{I}({\bf r},t)=\frac 1V\sum_{\bf q,k}e^{-i{\bf k\cdot r}}b^\dag_{{\bf
q}+ {\bf k}/2}b_{{\bf q}-{\bf k}/2}.
\end{equation}
The characteristic sizes of spatial inhomogeneities described by both operators, $ \hat{f}({\bf r},{\bf q},t)$ and $ \hat{I}({\bf r},t)$, are assumed much greater than the optical wavelength $\lambda=(2\pi/q_0)$. Here $q_0$ is the wave vector corresponding to the central frequency of the radiation, $\omega _0=cq_0$. Therefore it is quite reasonable to restrict the sums in Eqs. (\ref{1threee}) and (\ref{2my}) by the range of small $k$, i.e. $k< k_0$ such that the inequality $k_0\ll q_0$ is satisfied.  At the same time, the value of $k_0$ must be large enough to provide the  desired   spatial accuracy of the beam description.  If the characteristic size of the radiation inhomogeneity is $\delta r$, then the inequality   $k_0\delta r\gg1$ must be satisfied. Taking into account that the uncertainty of  ${\bf q}$ ($\delta q$) is  given by $(k/2)\le k_0/2$, we  conclude that  $\delta q \delta r\gg1/2$, satisfying the Heisenberg  inequality. The above considerations explain what types of optical properties can be studied using the distribution function. A similar idea of coarse-grained description of light has been used before (see, for example, the monograph by Mandel and Wolf \cite{Mandel} and Kolobov's review \cite{Kolobov}).

The operator of photon density can be represented as 
\begin{equation}\label{3my}
\hat{I} ({\bf r},t)=\langle\hat{I} ({\bf r},t)\rangle+\delta\hat{I} ({\bf r},t),
\end{equation}
where $ \langle\hat{I} ({\bf r},t)\rangle $ and $ \delta\hat{I} ({\bf r},t) $ are the average and fluctuating constituents of $ \hat{I} ({\bf r},t )$. The averaging includes the quantum-mechanical averaging of operators $\hat{f}$ and $\hat{I}$ and averaging over many ''runs" of the beam through different realizations of turbulence relief. Both averaging are performed independently.

Average product of photon densities is defined by
\begin{eqnarray}\label{4my}
\langle\hat{I}({\bf r}, t)\hat{I}({\bf r'}, t)\rangle{=}\hspace{50 mm}\\
\frac{1}{V^2}\sum_{\substack{{\bf q,k} \\ {\bf q',k'}}}e^{-i({\bf k}\cdot{\bf r}{+}{\bf k'}\cdot{\bf r'})}\langle b^\dag_{{\bf q}{+}\frac{\bf k}{2}}b_{{\bf q}{-}\frac{\bf k}{2}}b^\dag_{{\bf q'}{+}\frac{\bf k'}{2}}b_{{\bf q'}{-}\frac{\bf k'}{2}}\rangle .\nonumber
\end{eqnarray} 
All operators on the right side are given at time $ t $ which is a propagation time. It is reasonable to consider that the average product of four operators  at asymptotically large $ t $ splits into products of pairs, i.e., 
\begin{eqnarray}\label{5my}
\langle b^\dag_{{\bf q}{+}\frac{\bf k}{2}}b_{{\bf q}{-}\frac{\bf k}{2}}b^\dag_{{\bf q'}{+}\frac{\bf k'}{2}}b_{{\bf q'}{-}\frac{\bf k'}{2}}\rangle{\approx}\langle b _{{\bf q}{+}\frac{\bf k}{2}}^\dag b _{{\bf q}{-}\frac{\bf k}{2}}\rangle\langle b _{{\bf q}^\prime{+}\frac{{\bf k}^\prime}{2}}^\dag
b _{{\bf q}^\prime{-}\frac{{\bf k}^\prime}{2}}\rangle\hspace{5 mm}\\
{+}\langle b _{{\bf q}+\frac{\bf k}{2}}^\dag b _{{\bf q}^\prime -\frac{{\bf k}^\prime}{2}}\rangle\langle b _{{\bf q}^\prime+\frac{{\bf k}^\prime}{2}}^\dag
b _{{\bf q}{-}\frac{\bf k}{2}}\rangle{+}\langle b _{{\bf q}{+}\frac{\bf k}{2}}^\dag
b _{{\bf q'}{-}\frac{\bf k'}{2}}\rangle\delta_{{\bf q}{-}\frac{\bf k}{2}, {\bf q'}{+}\frac{\bf k'}{2}}.\nonumber
\end{eqnarray} 
In other words, it is assumed that amplitudes $b^\dag$ and $b$ obey Gaussian  statistics. Such assumption is justified since for $t\rightarrow\infty$ each primary coherent electromagnetic wave experiences multiple scatterings by randomly distributed vortices \cite{Das,DeWolf}. The "collisions" randomize photon momenta.  (This is in contrast to the paper \cite{Zhang} in which the Gaussian statistics  was assumed from the very beginning of photon trajectory.)  It is also considered that the condition of paraxial approximation is still fulfilled, and therefore the wave vector component $ q_z $ remains almost unchanged (approximately equal to  $ q_0 $). This means that the RI variation in the $z$-direction has no significant effect on propagation of paraxial beams \cite{Tatarskii1}.  This phenomenon is interpreted in \cite{baskov2018} as arising from relativistic length contraction of moving objects (the relative motion of vortices and photons). The perpendicular constituent $ {\bf q}_\bot  $, where $ {\bf q}_\bot =\{q_x,q_y\}$, grows with time like   Brownian particles moving in the $ {\bf q}_\bot $-space (see \cite{berm}). 

Substituting (\ref{5my}) into (\ref{4my}), we obtain   

\begin{eqnarray}\label{6my}
\langle\hat{I}({\bf r}, t)\hat{I}({\bf r'}, t)\rangle{\approx}\frac{1}{V^2}\sum_{\substack{{\bf q,k} \\ {\bf q',k'}}}e^{-i({\bf k}\cdot{\bf r}{+}{\bf k'}\cdot{\bf r'})}\times\hspace{25 mm}\\
\hspace*{-\abovedisplayskip}\left[\langle b _{{\bf q}+\frac{\bf k}{2}}^\dag b _{{\bf q}-\frac{\bf k}{2}}\rangle\langle b _{{\bf q}^\prime +\frac{{\bf k}^\prime}{2}}^\dag
b _{{\bf q}^\prime -\frac{{\bf k}^\prime}{2}}\rangle\right.{+}\langle b _{{\bf q}+\frac{\bf k}{2}}^\dag b _{{\bf q}^\prime -\frac{{\bf k}^\prime}{2}}\rangle\langle b _{{\bf q}^\prime+\frac{{\bf k}^\prime}{2}}^\dag b _{{\bf q}-\frac{\bf k}{2}}\rangle\nonumber\\
{+}\left.\langle b _{{\bf q}+\frac{\bf k}{2}}^\dag b _{{\bf q'}-\frac{\bf k'}{2}}\rangle\delta_{{\bf q}-\frac{\bf k}{2}, {\bf q'}+\frac{\bf k'}{2}}\right] .\nonumber\hspace{5 mm}
\end{eqnarray}
Utilizing here the definitions (\ref{1threee}) and  (\ref{2my}), we arrive at more meaningful form 
\begin{eqnarray}\label{7my}
\langle\hat{I}({\bf r}, t)\hat{I}({\bf r'}, t)\rangle=\delta({\bf r}-{\bf r'})\langle\hat{I}({\bf r}, t)\rangle+\langle\hat{I}({\bf r}, t)\rangle\langle\hat{I}({\bf r'}, t)\rangle\hspace{5 mm}\\
+\sum_{{\bf q},{\bf q'}}\langle f(\frac{{\bf r}+{\bf r'}}{2},{\bf q})\rangle\langle f(\frac{{\bf r}+{\bf r'}}{2},{\bf q'})\rangle e^{i({\bf q'}-{\bf q})\cdot({\bf r}-{\bf r'})}.\nonumber
\end{eqnarray}
The third term in the right side  is obtained by formal changing the indices in the operators $b_{\bf q}$, $b_{\bf q}^\dag$, and exponential multiplier as follows
\[
{\bf k}\rightarrow{\bf q}-{\bf q'}+\frac{{\bf k}+{\bf k'}}{2},\quad {\bf k'}\rightarrow{\bf q'}-{\bf q}+\frac{{\bf k}+{\bf k'}}{2}\tag{\ref{7my}a}\label{7.1my}\]
\[{\bf q}\rightarrow\frac{1}{2}\left({\bf q}+{\bf q'}+\frac{{\bf k}-{\bf k'}}{2}\right),\quad {\bf q'}\rightarrow\frac{1}{2}\left({\bf q}+{\bf q'}-\frac{{\bf k}-{\bf k'}}{2}\right).\]
It can be written in a simpler form
\[\bigg|\sum_{\bf q}\langle f\left(\frac{{\bf r}+{\bf r^\prime}}{2},{\bf q}\right)\rangle e^{i{\bf q}\cdot({\bf r^\prime}-{\bf r})}\bigg|^2\equiv \bigg| \mathscr{F}\left(\frac{{\bf r}+{\bf r'}}{2},{\bf r}-{\bf r'}\right)\bigg|^2,\tag{\ref{7my}b}\label{7myB}\] 
where  
\[
\mathscr{F}({\bf r}_1,{\bf r}_2)=\sum_{\bf q}\langle\hat{f}({\bf r}_1,{\bf q})\rangle\exp(-i{\bf q}\cdot{\bf r}_2)\tag{\ref{7my}c}\label{7myC}
\]
is the Fourier transform of $\langle\hat{f}({\bf r}_1,{\bf q})\rangle$  in {\bf q} domain.

The equation (\ref{7my}) represents the result of mixing four light waves  after propagation of the beam over long distances in the $z$ direction.
The first term in the right side describes the shot noise. It has a quantum nature and is important in the case of low photon density. 
The second term is the product of average photon densities. It does not consider the correlation  of light intensity at different positions ${\bf r}$ and ${\bf r^\prime}$  and is not present in the average product of fluctuations  $\langle\delta\hat{I}({\bf r}, t)\delta\hat{I}({\bf r^\prime}, t)\rangle $.
The third term, like the first one, describes fluctuations. It is given by a quadratic form of PDF  which in general case does not reduce to the product of photon densities  $\langle\hat{I}({\bf r}, t\rangle$.

The point $ {\bf r}= {\bf r}^\prime$ is the exception where the second and third terms become equal to each other. As a result, the scintillation index, $ \sigma^2 $, defined by
\begin{equation}\label{8my}
\sigma^2=\frac{{\langle\hat{I}^2}\rangle-{\langle\hat{I}\rangle^2}}{{\langle\hat{I}\rangle^2}},
\end{equation}
is equal to unity regardless of $\bf{r}$. [We neglect the shot noise here.] This is the well-known result for saturated regime of scintillation \cite{DeWolf}. It should be emphasized that this result is obtained only from the conditions of the Gaussian  statistics  for field operators in PDF.

The individual terms in Eq. (\ref{7my}) 
\begin{equation}\label{9my}
\langle f(\frac{{\bf r}+{\bf r'}}{2},{\bf q})\rangle \langle f(\frac{{\bf r}+{\bf r'}}{2},{\bf q'})\rangle e^{i({\bf q'}-{\bf q})\cdot({\bf r}-{\bf r'})}, 
\end{equation}
contribute to the correlation function of photon densities in different points ${\bf r}$ and ${\bf r'}$. At the same time, both photon distribution functions in (\ref{9my}) depend on the same spatial coordinate ${({\bf r}+{\bf r'})}/{2}$ which is different from ${\bf r}$ and ${\bf r'}$.
As a result the correlation of photon densities at different positions ${\bf r}$ and ${\bf r'}$ depends on the values of PDF at  $({\bf r}+{\bf r'})/2$. In particular, this may be the beam center (if ${\bf r}=-{\bf r'}$) where the PDF can have the maximum value. The discussed example indicates that by studying the photon  fluctuations outside the center of the beam, we can get information about its central region. 

The physical nature of the third term is similar to the Hanbury Brown-Twiss effect \cite{Zubairy} which describes four-wave correlations. The properties of the radiation field which are due to the spatial correlation of the irradiation density can be used in practice \cite{Garnier}. 

Explicit value for the third term is very simple  in "coordinate representation" (widely used in theory of photoelectric measurements, see \cite{Mandel}) which is based on modified light amplitudes. The latter are given by 
\[ b_{\bf r}=(b^\dag_{\bf r})^\dag=\frac{1}{\sqrt{V}}\sum_{\bf q}e^{i{\bf qr}}b_{\bf q}.\]
Using these variables, the third term is expressed as $|\langle b^{\dag}_{\bf r} b_{\bf r^\prime}\rangle|^2$. To obtain such a simple formula,  the definition of PDF  was used. 

It is reasonable to consider that for large values of difference $|{\bf r}-{\bf r}^\prime|$, there is no correlation between amplitudes $b_{\bf r}^\dag$ and $b_{{\bf r}^\prime}$. Conversely, non-zero value of $|\langle b^{\dag}_{\bf r} b_{\bf r^\prime}\rangle|^2$ indicates the presence of correlation of photon density at the points ${\bf r}$ and ${\bf r^\prime}$. Using two detectors at these points, we can directly obtain the correlation function of fluctuations and compare it with the theoretical value $|\langle b^{\dag}_{\bf r} b_{\bf r^\prime}\rangle|^2$. 

Equation (\ref{8my}) represents a relative fluctuation of  photon density at the specified locations where ${\bf r}={\bf r}^\prime$. It is reasonable to expect that the total  photon flux in the $z$-direction  defined by 
\begin{equation}\label{10my}
I_T(z,t)=c\int d{\bf r}_\bot \hat{I}({\bf r})   
\end{equation}  
does not fluctuate due to the lack of absorption. The following considerations elucidate this point.  A size of fluctuations of $I_T(z,t)$ can be described by the average  $\langle [\delta I_T(z,t)]^2\rangle $. Integration in the $(x,y)$ plane results in 

\begin{equation}\label{11my}
\langle [\delta I_T(z,t)]^2\rangle =c^2S\sum_{{\bf q}_\bot}\int d{\bf r}_\bot \langle\hat{f}({\bf r},{\bf q})\rangle^2.   
\end{equation} 
Equation (\ref{11my}) is obtained  using the relationships 
\[\int d({\bf r}_\bot{-}{\bf r}_{\bot}^\prime )e^{i({\bf q}_\bot-{\bf q}_{\bot}^\prime)\cdot({\bf r}_\bot-{\bf r}_{\bot}^\prime)}=(2\pi)^2\delta({\bf q}_\bot{-}{\bf q}_{\bot}^\prime)\]
and\[\delta({\bf q}_\bot{-}{\bf q}_{\bot}^\prime)= \frac S{(2\pi)^2}\delta_{{\bf q}_\bot,{\bf q}_{\bot}^\prime}.\]
The estimate for the relative fluctuations, which uses Eq.(\ref{8my}), is given by 
\begin{equation}\label{12my}
\frac{\langle [\delta I_T(z,t)]^2\rangle}{{\langle  I_T(z,t)\rangle}^2}\propto ({\bar q}_\bot R)^{-2},
\end{equation}
where symbols ${\bar q}_\bot$ and $R$ denote characteristic momentum of $q_\bot$ and the beam radius, respectively. Both of these quantities increase with  time $t$. Therefore the right side of Eq. (\ref{12my}) asymptotically goes to zero. This is a typical  situation for many-particle systems: usually relative fluctuations of total numbers of particles are negligible. The same applies to multiphoton system in this work.

Behavior of physical quantities ${\bar q}_\bot$ and $R$ was  investigated in earlier papers (see, for example, \cite{berm}) using the approach of ``slowly varying force". In what follows, our consideration is based on the alternative collision-integral method.  

\section{Boltzmann-Langevin equation for photon distribution function.}
\label{sec:BLE} 

Kinetic equation governing the evolution of photon density in the phase space,  $\hat{f}({\bf r},{\bf q},t)$, can be represented by \cite{baskov2018}
\begin{equation}\label{13my}
\partial_t \hat{f}({\bf r},{\bf {q}},t)+{\bf c_q}\cdot\partial_{\bf r}\hat{f}({\bf r},{\bf q},t)=
\hat{K}({\bf r},{\bf q},t)-
\hat{\nu}_{\bf q}\big \{ \hat{f}({\bf r},{\bf q},t)\},
\end{equation}
where ${\bf c_q}= { \partial\omega_q}/{\partial\bf q}=c{\bf q}/q_0$. Random nature of individual collisions is  taken into account by the Langevin source of fluctuations, $\hat{K}({\bf r},{\bf q},t)$. The Langevin  approach is widely used in quantum optics \cite{Zubairy} and physics of solids \cite{Kogan}. This approach uses stochastic differential equation for modeling the dynamics of physical systems (see \cite{BHARUCH}). 
The ``collision" integral $\hat{\nu}_{\bf q}$ is given by
\begin{eqnarray}\label{14my}
\hat{\nu}_{\bf q}\big \{ \hat{f}({\bf r},{\bf q},t)\}=\frac{2\pi\omega_{0}^{2}}{c}&\int& d{\bf k'_{\bot}}\psi({\bf k'_{\bot}})\\
&\times&\big(\hat{f}({\bf r},{\bf q},t)-\hat{f}({\bf r},{\bf q+k'_{\bot}},t)\big),\nonumber
\end{eqnarray} 
where
$\psi ({\bf k'_{\bot}})=\frac V{(2\pi)^3}\langle|n_{\bf k'_{\bot}}|^2\rangle$ and  $n_{\bf k}$ is the Fourier transform of the fluctuating refraction index
$\delta n({\bf r})$:  
\begin{equation}\label{15my}
n_{\bf k}=\frac 1V\int dVe^{i{\bf k\cdot r}}\delta n({\bf r}).
\end{equation}
The von Karman formula   
\begin{equation}\label{16my}
\psi ({\bf k})=0.033C_n^2\frac {\exp(-(kl_0'
)^2)}{(k^2+L_0^{-2})^{11/6}},
\end{equation}
provides a reliable description of atmospheric turbulence by means of the set of parameters $C_n^2$, $l_0'$ and $L_0$.  These parameters take into account the turbulence strength and the vortex shape. 

Linear inhomogeneous equation (\ref{13my}) can be used for the description of both the evolution of photon distribution and photon fluctuations. The term $\hat{\nu}_{\bf q}\big \{ \hat{f}({\bf r},{ \bf q},t)\}$ describes photon-eddies scattering. It follows from Eq. (\ref{14my}) that the quantity
\[
W({\bf k'_{\bot}})=\frac{(2\pi)^3\omega_{0}^{2}}{Sc}\psi({\bf k'_{\bot}})
\]
can be interpreted as the probability  per unit time of a photon transition from state 
${\bf q}_{\bot}$  to state  ${\bf q}_{\bot}\pm{\bf k}'_{\bot} $. In what follows, we will use the quantity 
\begin{equation}\label{17my}
\nu=\frac{2\pi\omega_{0}^{2}}{c}\int d{\bf k'_{\bot}}\psi({\bf k'_{\bot}})
\end{equation}
as a convenient parameter which describes the relaxation in the system and does not depend on photon variables.

After averaging of Eq. (\ref{13my}), the Langevin source disappears and the kinetic equation for $ \langle \hat{f}({\bf r},{\bf q},t)\rangle$ reduces to  
\begin{equation}\label{18my}
[\partial_t+{\bf c_q}\cdot\partial_{\bf r}+\hat{\nu}_{\bf q}] \langle\hat{f}({\bf r},{\bf {q}},t)\rangle=0.
\end{equation}
The kinetic equation for the fluctuating component 
\begin{equation}\label{19my}
\delta f({\bf r},{\bf {q}},t)=\hat{f}({\bf r},{\bf {q}},t)-\langle\hat{f}({\bf r},{\bf {q}},t)\rangle
\end{equation}
is given by
\begin{equation}\label{20my}
[\partial_t+{\bf c_q}\cdot\partial_{\bf r}+\hat{\nu}_{\bf q}] \delta f({\bf r},{\bf {q}},t)=  \hat{K}({\bf r},{\bf q},t).
\end{equation}

Two linear equations (\ref{18my}) and (\ref{20my}) can be used for obtaining average and fluctuating PDF. The state of a radiation field in the source aperture determines the boundary conditions for $\langle\hat{f}\rangle $.  The free term, $ K $, distributed along the beam path, is the only source of non-trivial solution of (\ref{20my}) if photon fluctuations in the outgoing aperture  are neglected.    

\section{Analytical solution for average PDF.}
\label{sec:average_pdf}

The  homogeneous equation  (\ref{18my}) can be solved using the Fourier transforms. To do this, we multiply all terms in $(\ref{18my})$ by $\frac 1S\times e^{i{\bf k_\bot}\cdot{\bf r}}$ and  integrate over ${\bf r_\bot}$.  [In what follows, we use the notation $f({\bf r},{\bf q},t)$ for $\langle \hat{f}({\bf r},{\bf q},t)\rangle$.] As a result, we obtain for  the Fourier component, defined by 
\begin{equation}\label{21my}
f({\bf k_\bot},{\bf q},t)=\frac{1}{S}\int d{\bf r_\bot}f({\bf r},{\bf {q}}, t)e^{i{\bf k_\bot}\cdot{\bf r}},
\end{equation}
a simpler equation   
\begin{equation}\label{22my}
(\partial_t-i{\bf k}_\bot \cdot{\bf c}_{{\bf q}_\bot})f({\bf k_\bot},{\bf q_\bot},t)=-\hat{\nu}_{\bf q_\bot}\big \{ f({\bf k}_\bot,{\bf q}_\bot,t)\},
\end{equation}
in which ${\bf c}_{{\bf q}_\bot}=c{{\bf q}_\bot}/{q_0}$ and $ |{\bf c}_{{\bf q}_\bot}|\ll c$. In the course of derivation of Eq. (\ref{22my}) we
consider $z=ct$ (see Appendix \ref{appendix1}). Since  the $z$ components of  vectors are not  in the last equations,  we omit the notation $\bot$ for the  brevity of the body text (except Appendix \ref{appendix2})  and consider   only $x$- and $y$-components.
Solution of equation (\ref{22my}) can be represented as the product 
\[f({\bf k},{\bf q},t)=e^{i{\bf k}\cdot{\bf q}\frac{ct}{q_0}}\times\varphi({\bf k},{\bf q},t),\] where the new function $\varphi$ satisfies the integral equation 
\begin{eqnarray}\label{23my}
\partial_t\varphi({\bf k},{\bf q},t){=}{-}\frac{2\pi\omega_{0}^2}{c}{\int}d{\bf k'}\psi({\bf k'})\times\hspace{35 mm}\\
\bigg[\varphi({\bf k},{\bf q},t)
-e^{i{\bf k}\cdot{\bf k'}\frac{ct}{q_0}}\varphi({\bf k},{\bf q}{+}{\bf k'},t)\bigg].\nonumber
\end{eqnarray}
Another Fourier transform,  
\begin{equation}\label{24my}
\Phi({\bf k},{\bf p}, t)=S^{-1}\sum_{\bf q}e^{{-i\bf p}\cdot{\bf q}}\varphi({\bf k},{\bf q},t),
\end{equation}
reduces the kinetic equation to the relaxation type equation   
\begin{equation}\label{25my}
\partial_t\Phi({\bf k},{\bf p},t)=-\gamma({\bf k},{\bf p},t)\Phi({\bf k},{\bf p},t).
\end{equation}
The relaxation frequency $\gamma$ is given by
\begin{equation}\label{26my}
\gamma({\bf k},{\bf p},t)=\frac{4\pi\omega_{0}^{2}}{c}\int d{\bf k'_{\bot}}\psi({\bf k'_{\bot}})\sin^2\bigg[\bigg({\bf p}-{\bf k}\frac{ct}{q_0}\bigg){\cdot} 
\frac{\bf k'_{\bot}}2\bigg].
\end{equation} 
The set of independent variables ${{\bf p},\bf k},t$,  forms the vector ${\bf P}(t)={\bf p}-{\bf k} tc/q_0$,  on which  the relaxation parameter depends: $\gamma\equiv\gamma\big({\bf P}(t)\big)$. In the case of large characteristic values of  ${\bf P}(t)\cdot {\bf k}^\prime/2$, the quantity of $\gamma$ approaches $\nu$ [see Eq. (\ref{17my})].

Solution of Eq. (\ref{25my}) is given by
\begin{equation}\label{27my}
\Phi({\bf k},{\bf p},t)=e^{-\int_0^tdt^\prime \gamma({\bf k},{\bf p},t^\prime)}\Phi({\bf k},{\bf p},t=0).
\end{equation}
The initial value of $\Phi({\bf k},{\bf p},t=0)$ depends on the field in the source aperture, that is, when $z=ct=0$. It is assumed that there are no  fluctuating components in this field.   

Using Eq. (\ref{27my}), we obtain
\begin{eqnarray}\label{28my} 
f({\bf k},{\bf q},t){=}\int d{\bf p}e^{-i{\bf q}\cdot({\bf p}-{\bf k} ct/q_0)}e^{-\int_0^t\gamma({\bf k},{\bf p},t^\prime)}\Phi({\bf k},{\bf p},t{=}0).\nonumber\\
\end{eqnarray}
General solution (\ref{28my}) is analogous for the result in \cite{Manning} (see Eq. (4.2) there) dealing with radiative transfer theory for inhomogeneous media. Although the paper considers direction-dependent radiance distribution function $I({\bf R}, {\bf n})$(notations as in \cite{Manning}) it could be viewed as classical counterpart for  $\hat{f}({\bf r},{\bf q},t)$. 

In the case of a Gaussian  profile of the source field described by function $ e^{-r^2/r_0^2}$,  the initial value of ${\varphi ({\bf k},{\bf q},t=0)}$  is proportional to $\exp\left(-q^2 \frac {r_0^2}2-k^2 \frac {r_0^2}8\right)$ (see, for example, Ref. \cite{baskov2016}). Then the average photon distribution function is given by 
\begin{eqnarray}\label{29my}
f({\bf r},{\bf q},t){=}C\int d{\bf p}\int d{\bf k}e^{-i{\bf k} \cdot{\bf r}-i{\bf q}\cdot({\bf p}-{\bf k} ct/q_o)}\nonumber\\
\times e^{- k^2r_0^2/8-p^2/2r_0^2}e^{-\int\limits_0^tdt'\gamma({\bf k},{\bf p},t')},
\end{eqnarray}
where the constant $C$ can be  expressed in terms of the total photon flux. When deriving Eq. (\ref{29my}),  the relations (\ref{24my}) and (\ref{28my}) were taken into account. 

The term (\ref{29my}) is an analytic solution in the paraxial approximation of the kinetic equation (\ref{18my}). Finding the average distribution function  requires multiple  (sevenfold) integration  on the right side of Eq. (\ref{29my}).  However, some qualitative conclusions about the shape of PDF are possible without numerical integration. First of all, we consider the case of  long propagation time when the characteristic values of $k^2$ and ${\bf p}$ are small. The reasons for their decreasing are: (i) beam broadening and (ii) increase of photonic momenta ${\bf q} $ due to its random (Brownian) dynamics. As a  result, the average PDF is simplified to  
\begin{eqnarray}\label{30my}
f({\bf r},{\bf q},t){=}C\int d{\bf p}\int d{\bf k}e^{-i{\bf k} \cdot({\bf r}-{\bf q} ct/q_0)}e^{-i{\bf q}\cdot{\bf p}}e^{-\int\limits_0^tdt'\gamma({\bf k},{\bf p},t')}.\nonumber\\
\end{eqnarray}
The first multiplier of the integrand is an oscillating function. However, it is equal to unity if ${\bf r}={\bf q} ct/q_0$. As a result, $f({\bf r},{\bf q},t)$ is larger if vectors $\bf r$ and ${\bf q}$ are  equally oriented. Drift of photons with the velocity ${\bf q} c/q_0$ is responsible for the asymmetry of their distribution in the (${\bf r} ,{\bf q}$) space.

Distribution functions  depend  on the parameters of  source radiation and atmospheric turbulence.  
Further consideration is simplified if the quantity $L_0^{-2}$ in the denominator of Eq. (\ref{16my}) is  omitted. This simplification changes the expression for $C_n^2$  to the Tatarskii version, which is often used in the atmospheric optics. Then we can  integrate in (\ref{26my}) over ${\bf k}^\prime$ and find an analytic expression for the relaxation frequency $\gamma$. It is given by

\begin{eqnarray}\label{31my}
\gamma({\bf k},{\bf p},t){=}\frac{2\pi^2\omega_{0}^{2}}{c}0.033C_n^2\Gamma\left(-\frac{5}{6}\right){l_0'}^{5/3}\nonumber\\
\times\left[1-{}_1 F_1\left(-\frac{5}{6},1;-\frac{({\bf p}-{\bf k}\frac{ct}{q_0})^2}{4{l_0'}^2}\right)\right]
\end{eqnarray}
where ${}_1 F_1(a,b;z)=\sum\limits_{n=0}^{\infty}\frac{a^{(n)}z^n}{b^{(n)}n!}$ is a confluent hypergeometric function (Kummer's  function) and  $a^{(n)}$, $b^{(n)}$ are the Pochhammer symbols. One might see that solution (\ref{28my}) with relaxation frequency (\ref{31my}), having it summarized over ${\bf q}$, agrees with results for mutual coherence function in earlier works \cite{Tatarskii1,Fante1,Andriews3}.


\subsection{Transverse momenta of photons}
The mean square of the transverse momentum is defined by
\begin{equation}\label{32my}
\langle{\bf q}^2\rangle=\frac{\int d{\bf r}\int d{\bf q}q^2 f({\bf r},{\bf q},t)}{\int d{\bf r}\int d{\bf q}f({\bf r},{\bf q},t)}.
\end{equation}
It should be reminded that all vectors have only perpenpicular to $z$ components. The integration over ${\bf r}$ is readily simplified since it considers only factor $e^{-i{\bf k}\cdot{\bf r}}$ in $f({\bf r}, {\bf q},t)$: $\int d{\bf r}e^{-i{\bf k}\cdot{\bf r}}=(2\pi)^2\delta({\bf k})$. As before, we consider  the initial profile of the laser beam to be of a Gaussian form. Using Eqs. (\ref{29my}) and integrating over ${\bf r}$ in  (\ref{32my}), we obtain  
\begin{equation}\label{33my}
\langle{\bf q}^2\rangle=\frac{\int d{\bf q}q^2\int d{\bf p} e^{-i{\bf p}\cdot{\bf q}} e^{-p^2/2r_0^2}e^{-\gamma_p t}}{\int d{\bf q}\int d{\bf p}e^{-i{\bf p}\cdot{\bf q}} e^{-p^2/2r_0^2}e^{-\gamma_pt}},
\end{equation}
where $\gamma_p\equiv\gamma({\bf k}=0,{\bf p},t=0)$ is given by 
\begin{equation}\label{34my}
\gamma_p=\frac{4\pi\omega_{0}^{2}}{c}\int d{\bf k'_{\bot}}\psi({\bf k'_{\bot}})\sin^2(\frac 12 {\bf p}{\cdot} 
{\bf k'_{\bot}}).
\end{equation} 
For long-distance propagation (large $t$), only small values of $\gamma_p$ make a significant contribution to (\ref{33my}). In this case, we can approximate $\sin^2( \frac 12{\bf p}{\cdot}{\bf k'_{\bot}})$  by the value 
$\frac 14({\bf p}{\cdot}{\bf k'_{\bot}})^2$. 
Then the variables $ {\bf k}^\prime $ and $ {\bf p} $ in the expression for $\gamma_p $ are separated from each other resulting in
\begin{equation}\label{35my}
\int d{\bf p} e^{-i{\bf p}\cdot{\bf q}} e^{{-p^2/2r_0^2}{-\gamma_p t}}=\int d{\bf p} e^{-i{{\bf p}\cdot{\bf q}} - p^2(\alpha t+1/2r_0^2)},  
\end{equation}
where a constant $\alpha$ is given by
\begin{equation}\label{36my}
\alpha =\frac{\pi\omega_0^2}{2c}\int d{\bf k}^\prime  k^{\prime 2}\psi(k^\prime) .
\end{equation}
After integration in (\ref{33my}), we obtain 
\begin{equation}\label{37my}
\langle{\bf q}^2\rangle=\frac2{r_0^2}+4\alpha t  ,
\end{equation} 
and finally
\begin{equation}\label{38my}
\langle{\bf q}^2\rangle=\frac{2}{r_0^2}+0.066\pi^2
\Gamma\left(\frac{1}{6}\right)
C_n^2q_0^2{l_0^{'}}^{-\frac{1}{3}}z
\end{equation}
for Tatarskii spectrum. It can be seen from Eq. (\ref{35my}) that the main contribution to the integral is given by the region where the condition $t\gamma_p\sim1$ holds. The estimate of this quantity for large $t$ is given by  $ (k^\prime p)^2 \nu t$  where  $ k^\prime$ and $p^2$ are the corresponding characteristic values. Taking into account that $(k^\prime p)^2\sim (k^\prime/ q)^2\ll1$, we find that
\begin{equation}\label{39my}
\nu t\gg1.
\end{equation}
This inequality  means the propagation time $t$ should be much longer than the relaxation time $\nu^{-1}$. It is the general condition for  validity of the asymptotics (\ref{38my}).   The mean square of the transverse momentum was usually described by coherence length, denoted by $\rho_0$, (see earlier works \cite{Fante3, Churnside}). The result (\ref{37my}) is identical for values of $\rho_0$ for corresponding asymptotic conditions.

The first term in (\ref{37my}) is due to the diffraction of the beam at the source aperture. 
The second one is identical to that derived in \cite{berm} using a random-force approach [see Eq. (53) there]. The equivalence of results obtained  by  different methods is not occasional. This point can be explained as follows. Increasing the propagation time leads to a decrease in the ratio $ k^\prime/q$ due to the increase in photon momentum $q$. The physical picture based on ``photon-eddy collisions" becomes more like a picture in which the interaction of photons with atmosphere is described by a ``random force ${\bf F}$". Small value of  $ k^\prime/q$ provides smoothing  of this interaction.  The equivalence of these approaches at large propagation time  is shown in the Appendix \ref{appendix2}. 

There is  a  considerable interest in a region  close to the source aperture. A simple analysis  clarifies the initial stage of the growth of the momentum $\langle {\bf q}^2 \rangle $ (starting at $ 2/r_0^2) $. An  analytical expression that takes into account the linear dependence on $ t $ is easy to obtain. For small values of $t$, the exponent $e^{-\gamma_p t}$ can be  approximated by  $1-\gamma_p t$. Integration of  (\ref {33my}) over the variables 
$ {\bf p}, {\bf q}, {\bf k}^\prime $ leads to  
\begin{equation}\label{40my}
\langle{\bf q}^2\rangle=\frac{2}{r_0^2}+0.066\pi^2
\Gamma\left(\frac{1}{6}\right)
C_n^2q_0^2{l_0^{'}}^{-\frac{1}{3}}z.
\end{equation}
It can be seen that the expressions (\ref{38my}) and 
(\ref{40my}) are identical, though they are applicable for very different values of time $t$. It seems that the coincidence of Eqs. (\ref{38my}) and (\ref{40my}) is occasional. Equation (\ref{40my}) holds for condition
\begin{equation}\label{41my}
\nu t\ll1.
\end{equation}
The size of $\langle{\bf q}^2\rangle$ for any physical parameters, included in Eq. (\ref{28my}), can be calculated numerically using Eq. (\ref{31my}). 

\subsection{Beam broadening}

Beam broadening is characterized by the average value of the square of the distance from the $z$ axis. Following the scheme of the previous subsection,  we substitute $ r^2$ for $ q^2$ in Eq. (\ref{32my}) 
\begin{equation}\label{42my}
\langle{\bf r}^2\rangle=\frac{\int d{\bf r} d{\bf k} r^2 e^{-i{\bf k}\cdot {\bf r} -\frac {k^2r_0^2}8(1+\frac{4c^2t^2}{q^2_0r^4_0})-\int_0^tdt^\prime\gamma (t^\prime)}}{\int d{\bf r} d{\bf k}  e^{-i{\bf k}\cdot  {\bf r} -\frac {k^2r_0^2}8(1+\frac{4c^2t^2}{q^2_0r^4_0})-\int_0^tdt^\prime\gamma (t^\prime)}},
\end{equation}
where integration over variables  ${\bf q}$ and ${\bf p}$  has already been done. The relaxation term $\gamma (t^\prime)$ in Eq. (\ref{42my}) is given by
\begin{equation}\label{43my}
\gamma (t^\prime)=\frac{4\pi\omega_0^2}c\int d{\bf k}^\prime\psi (k^\prime) \sin^2\big({\bf k} \cdot {\bf k}^\prime\frac{ct^\prime}{2q_0}\big)
\end{equation}
As before, in the case of large times $t$, the sine in (\ref{43my}) can be replaced by its argument.  This simplifies integration into Eq. (\ref{42my}). The result is 
\begin{equation}\label{44my}
\langle{\bf r}^2\rangle=\frac{r_0^2}2\bigg( 1+\frac {4z^2}{r_0^4q_0^2}+\frac{8z^3c\alpha}{3r_0^2\omega_0^2} \bigg).
\end{equation}
Expression for beam broadening (\ref{44my}) reproduces the result of earlier approaches \cite{Andrews2, berm, berm2009} for long-distance propagation.
It can be seen from Eq. (\ref{44my}) that the last term  in the parentheses describes long-distance  behavior ($\propto z^3$) of the beam radius.

\subsection{Asymptotic distribution function}

%
An analytical expression for the long-distance value of PDF can be easily obtained after integrating over ${\bf p}$ and  ${\bf k}$ in Eq. (\ref{30my}). The result is 
\begin{eqnarray}\label{45my}
f({\bf r},{\bf q},t){=}3C\bigg(\frac{2\pi q_0}{ct^2\alpha}\bigg)^2\exp{\left[{-}\left({\bf r}{-}\frac{{\bf q}ct}{2q_0}\right)^2\frac 4{\langle {\bf r}^2\rangle_T}{-}\frac{4{q}^2}{\langle {\bf q}^2\rangle_T}\right]}\nonumber\\
\end{eqnarray}
where $\langle {\bf r}^2\rangle_T={4z^3c\alpha}/({3\omega_0^2})$ and $\langle {\bf q}^2\rangle_T=4\alpha t$ describe beam broadening and increase of photon momentum  caused by atmospheric turbulence [see Eqs. (\ref{37my}) and (\ref{44my})]. (As previously, the constant $C$ is expressed by the total photon flux in the $z$ direction.) For large $t$, the form of $f$ is independent of the initial (at $t=0$) configuration of the radiation field.

\begin{figure}[ht]
\centering
\includegraphics[width=0.98\linewidth,keepaspectratio]{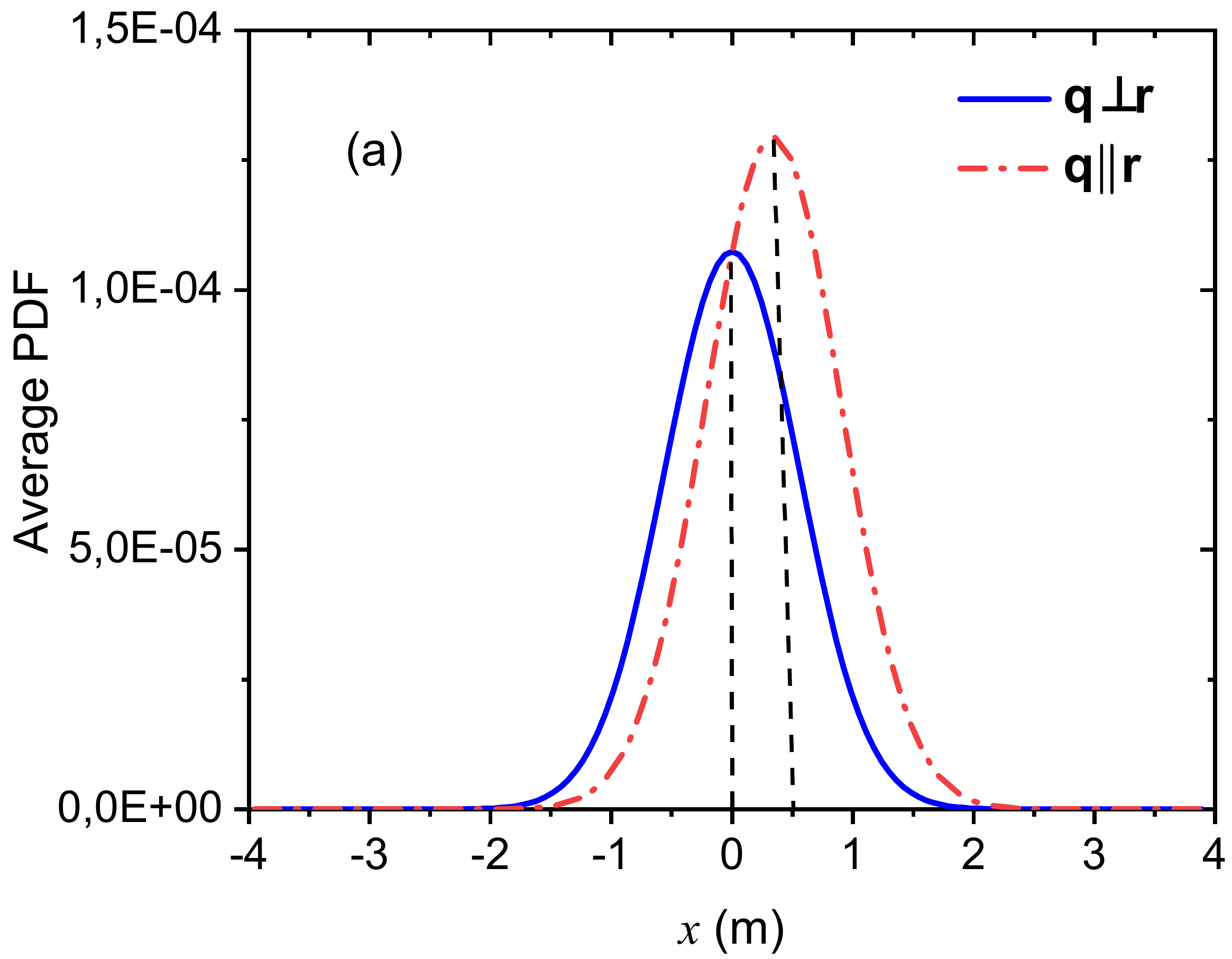}
\includegraphics[width=0.98\linewidth,keepaspectratio]{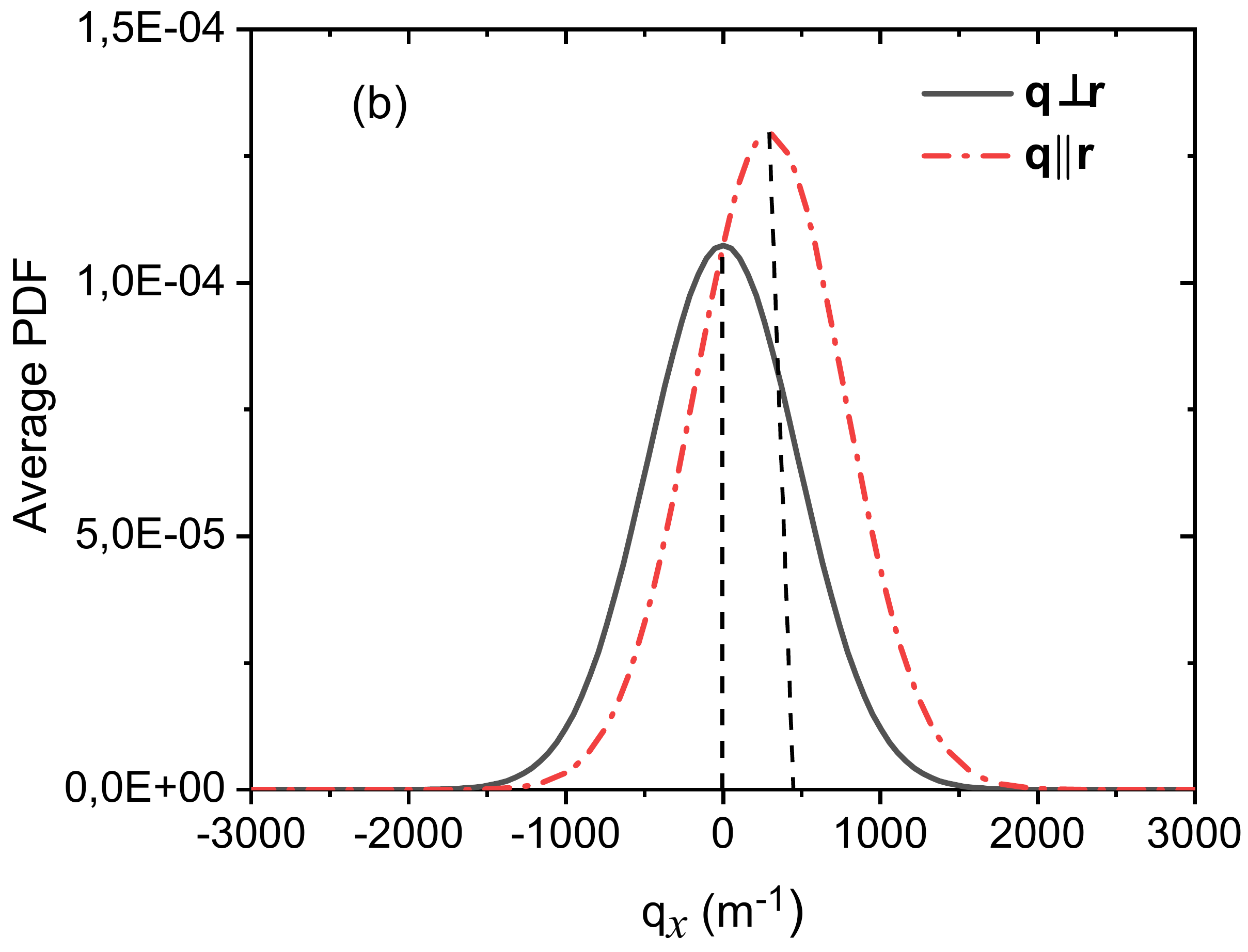}

\caption{ PDF profile in (a) ${\bf r}$-space with fixed $y=0$,  $q_x=0$, $q_y=\sqrt{\alpha t}\equiv\sqrt{\langle {\bf q}^2\rangle_T}/2$ for solid curve, and $q_x=\sqrt{\langle {\bf q}^2\rangle_T}/2$, $q_y=0$ for dash-dotted curve; (b) ${\bf q}$-space with fixed  $q_y=0$,  $x=0$, $y=\sqrt{\frac{z^3\alpha}{3q_0^2c}}\equiv\sqrt{\langle {\bf r}^2\rangle_T}/2$ for solid curve, and $x=\sqrt{\langle {\bf r}^2\rangle_T}/2$, $y=0$ for dash-dotted curve. Parameters of the beam and channel: $r_0=0.01\,m$, $z=20\,km$, $C_n^2=2.5\times10^{-14}\,m^{-2/3}$, $l_0/2\pi=10^{-3}\,m$, $q_0=10^{7}\,m^{-1}$.}
\label{fig:zero}
\end{figure}

Spatial distribution is symmetric with respect to the direction of ${\bf q}$ and reaches maximum value when ${\bf r}=\frac{{\bf q}ct}{2q_0}$ (see Fig. \ref{fig:zero}). After summing over ${\bf q}$ both sides of the
Eq. (\ref{45my}), we obtain the asymptotic value of the photon density, 
\begin{equation}\label{46my}
\langle \hat{I}({\bf r},t)\rangle=C \frac{4\pi S}{\langle{\bf r}^2\rangle_{T}}e^{-\frac {r ^2}{\langle{\bf r}^2\rangle_{T}}},
\end{equation}
which represents the second moment of the radiation field. Then the shot noise, given by the first term in the right side of Eq. (\ref{7my}), reduces to
\begin{equation}\label{47my}
\langle {\delta I}^2_{shot}\rangle=C \frac{4\pi S}{\langle{\bf r}^2\rangle_{T}}\delta ({\bf r}-{\bf r^\prime})e^{-\frac {r ^2}{\langle{\bf r}^2\rangle_{T}}}=\langle I({\bf r},t)\rangle\delta ({\bf r}-{\bf r^\prime}),
\end{equation}
where the paraxial approximation was assumed.
Equation (\ref{47my}) is  the Poissonian-like component of the fourth-order  moment derived in Sec. \ref{sec:pdf}. In the next Section, the  
contribution of the shot noise to the full noise is analyzed and interpreted.    

\section{Asymmptotic fourth moment and aperture averaging of fluctuations}
\label{sec:forth_moment}

The full expression for the fourth moment is obtained after integration of  Eq. (\ref{46my}) over ${\bf q}$ and ${\bf q}^\prime$. As a result, we get
\begin{eqnarray}\label{48my}
&&\langle\hat{I}({\bf r}, t)\hat{I}({\bf r'}, t)\rangle=\langle {\delta I}^2_{shot}\rangle+\langle\hat{I}({\bf r}, t)\rangle\langle\hat{I}({\bf r'}, t)\rangle\nonumber+\\
&&\left(\frac{C4\pi S}{\langle{\bf r}^2\rangle_{T}}\right)^2\exp\left\{{-}\frac{({\bf r}{+}{\bf r'})^2}{2\langle{\bf r}^2\rangle_{T}}{-}\frac{({\bf r}{-}{\bf r'})^2\langle{\bf q}^2\rangle_{T}}{8}\right\},
\end{eqnarray}
where the distribution function (\ref{45my}) was used. It is seen from the equation (\ref{48my}) that the characteristic transverse momentum determines the correlation length. This length is of the order of $(8/{{\langle{\bf q}^2\rangle}_{T}})^{1/2}$ and  decreases with the propagation time as $t^{-1/2}$. Loss of the coherence means that different portions of the radiation field behave independently of each other as non-interacting particles. The long-time asymptotics for the exponential factor in Eq. (\ref{48my}) tends to delta function, the argument of which is the difference of spatial variables [$\sim \delta ({\bf r}-{\bf r^\prime})$]. Then the last term in (\ref{48my}) can be approximated by   
\begin{equation}\label{49my}
\delta({\bf r}-{\bf r}^\prime)\frac{8\pi}{\langle{\bf q}^2\rangle_{T}}\langle I({\bf r},t)\rangle^2 ,
\end{equation}
where the relation (\ref{46my}) was used. We can now express the full term for the fluctuations of the photon density as
\begin{equation}\label{49myy}
\langle\delta\hat{I}({\bf r}, t)\delta\hat{I}({\bf r'}, t)\rangle=\delta({\bf r}-{\bf r}^\prime)\langle{\hat I}({\bf r},t)\rangle\bigg(1+\frac{8\pi}{\langle{\bf q}^2\rangle_{T}}\langle{\hat I}({\bf r},t)\rangle\bigg).
\end{equation}
The right side is the sum of the shot noise and the classical noise. The last is quadratic in the radiation density. In the case of high photon density, as one would expect, the classical noise prevails. However, photon density decreases with the increase of distance from the center of the beam, $r$, so the classical component becomes equal to and even less than the shot noise value (\ref{47my}). It is seen from Eq. (\ref{49myy})  that the equality occurs for some specific distance $r=r_{q}$ where  
\begin{equation}\label{50myy}
\langle I({r_q},t)\rangle=\frac{\langle{\bf q}^2\rangle_{T}}{8\pi}.
\end{equation}
The radius ${r_q}$ marks the boundary between areas with predominantly "classical" and "quantum" fluctuations of light.

The condition (\ref{50myy}) can be easily interpreted. The left-hand side describes the two-dimensional density of particles ("photons"). It can be expressed  as $\langle I({r_q},t)\rangle=a^{-2}$, where $a^2$ is a square per particle. On the other hand, the quantity $\langle{\bf q}^2\rangle_{T}$ in the right side is equal to $(2\pi/\lambda_q)^2$, where $\lambda_q$ is the characteristic wave length of particles in the two-dimensional domain. Using new notations, the condition (\ref{50myy}) can be rewritten in a more meaningful form:      
\begin{equation}\label{51myy}
\lambda_q=\sqrt{\frac {\pi}2}a. 
\end{equation}
The value on the left depends on time as $t^{-1/2}$. This means that  the increase of propagation time (or propagation distance ${z=ct}$) results in a similar decreasing the size of the "classical region" $a$. It is possible that for large $t$ the intensity of fluctuations acquires a completely quantum nature at any point of the beam cross-section. The physical picture is opposite to the phenomenon of the Bose condensation. In the process of propagation, classical waves becomes similar to an ensemble of free particles, between which there is no quantum correlation and which can be identified as a two-dimensional gas of non-interacting photons. In other words, the light beam is transformed into the flux of "free particles". 

At the end of the paragraph, it should be noted that for large (but still finite) values of $ \langle{\bf q^2}\rangle_{T} $ and small radius of receiving aperture, the $\delta $-function approximation for correlations does not apply. In this case, Eq. (\ref{48my}) should be used. This point is important when the averaging of $\Gamma_4$ [see next subsection] occurs in small areas of the $(x,y)$ plane. 

\subsection*{Aperture averaging}
The asymptotic value for fourth moment can be used to obtain the correlation function  of photon density fluctuations averaged over aperture plane of the detector. This physical value is important for measurements because the size of the receiver aperture is always limited \cite{vasylyev,Barrios,Vetelino}. To demonstrate the effect of the receiver aperture $\mathcal{A}$, we utilize intensity transmittance defined by
\begin{equation}
\eta=(4C\pi^2S)^{-1}\int\limits_{\mathcal{A}}d{\bf r} \hat{I}({\bf r},t),
\end{equation}
where we use normalizing condition $\eta =1$ for $\mathcal{A}$ which is much larger than the beam cross section. The quantity $\eta$ contains both the average and fluctuating constituents and describes transmittance of the optical channel \cite{semenov2012}. 

Fluctuations of the transmitted radiation are described by the aperture-averaged scintillation index    
\begin{equation}\label{scinteta}
\sigma_{\eta}^2=\frac{\langle\eta^2\rangle-\langle\eta\rangle^2}{\langle\eta\rangle^2}.
\end{equation}
Two moments  for $\eta$ are defined as \cite{semenov}
\begin{eqnarray}\label{eta_avg}
\langle\eta\rangle&=&\int\limits_{\mathcal{A}}d{\bf r} \Gamma_2({\bf r}),\\
\label{eta2_general}
\langle\eta^2\rangle&=&\int\limits_{\mathcal{A}}d{\bf r} \int\limits_{\mathcal{A}}d{\bf r}^\prime \Gamma_4({\bf r},{\bf r}^\prime),
\end{eqnarray}
where the normalizing conditions are given by relations: $\Gamma_2({\bf r})=\langle \hat{I}({\bf r},t)\rangle/(4C\pi^2S)$, $\Gamma_4({\bf r},{\bf r}^\prime)=\langle\hat{I}({\bf r}, t)\hat{I}({\bf r'}, t)\rangle/(4C\pi^2S)^2$.

For the round receiver aperture with radius $R$ and long-distance propagation we can express moments of transmittance as
\begin{eqnarray}
\langle\eta\rangle&=&\left(1-\exp\left(-\frac{R^2}{\langle{\bf r}^2\rangle_{T}}\right)\right),
\end{eqnarray}
\vspace*{-\abovedisplayskip}
\begin{eqnarray}\label{etha2_num}
\langle\eta^2\rangle{-}\langle\eta\rangle^2{=}\int\limits_0^{R^2}\frac{d\chi}{\langle{\bf r}^2\rangle_{T}} \int\limits_0^{R^2}\frac{d\chi^\prime}{\langle{\bf r}^2\rangle_{T}}I_0\left(\sqrt{\chi\chi^\prime}\left(\frac{1}{\langle{\bf r}^2\rangle_{T}}{-}\frac{\langle{\bf q}^2\rangle_{T}}{4}\right)\right)\nonumber\\
\exp\left(-(\chi{+}\chi^\prime)\left(\frac{1}{2\langle{\bf r}^2\rangle_{T}}{+}\frac{\langle{\bf q}^2\rangle_{T}}{8}\right)\right),\hspace{10 mm}
\end{eqnarray}
where $I_0$ is modified Bessel function of the first kind. The integral over $\chi^\prime$ in Eq. (\ref{etha2_num}) reduces to the incomplete cylindrical function, the analytical approximation of which has been studied in \cite{vasylyev2013}. 

\begin{figure}[ht]
\centering
\includegraphics[width=0.98\linewidth,keepaspectratio]{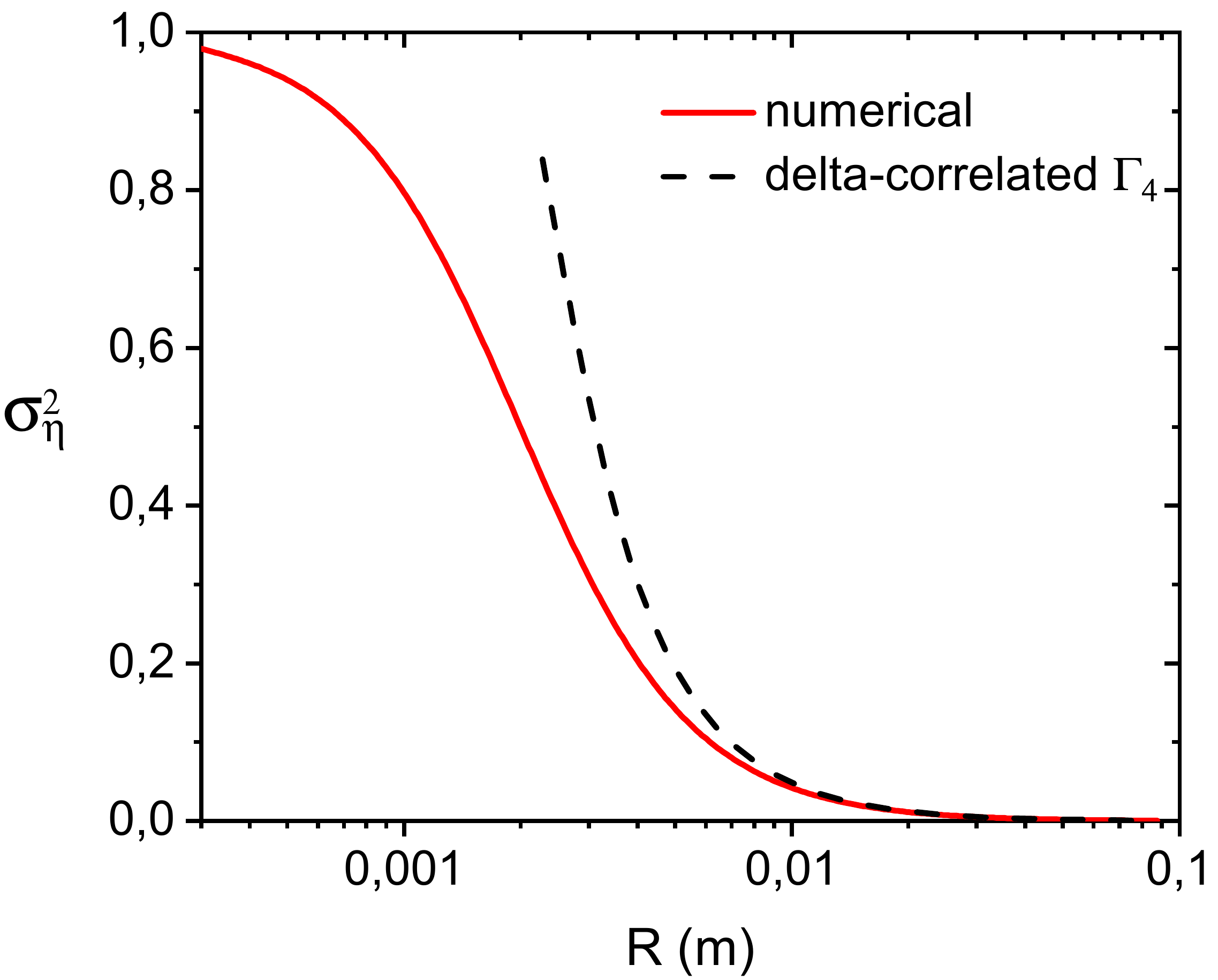}

\caption{ Aperture-averaged scintillation index vs. radius of receiver. Solid line represents the numerical results for  (\ref{etha2_num}); dashed line considers delta-correlated behavior of $\Gamma_4$, Eq. (\ref{etha2_delta}). Parameters of the beam and channel: $z=20\,km$, $C_n^2=2.5\times10^{-14}\,m^{-2/3}$, $r_0=0.01\,m$, $l_0/2\pi=10^{-3}\,m$, $q_0=10^{7}\,m^{-1}$. }
\label{fig:first}
\end{figure}

In the case of large $R$, the delta-correlation approximation (\ref{49my}) can be used. This considerably simplifies the calculation of (\ref{eta2_general}). The result is given by
\begin{equation}\label{etha2_delta}
\langle\eta^2\rangle{-}\langle\eta\rangle^2=\frac{4}{\langle{\bf q}^2\rangle_{T}\langle{\bf r}^2\rangle_{T}}\left(1-\exp\left(-\frac{2R^2}{\langle{\bf r}^2\rangle_{T}}\right)\right).
\end{equation}
For the realistic set of parameters, two curves are shown in Fig. \ref{fig:first}.  Both curves agree well only for large values of $R$. In this case, the value of scintillation can be estimated as
\[\sigma_{\eta}^2\sim\frac{1}{\langle{\bf q}^2\rangle_{T}\langle{\bf r}^2\rangle_{T}}\approxeq (\langle{\bf r}^2\rangle\langle{\bf q}^2\rangle)^{-1}.\] The same estimate was obtained using the qualitative analysis in Sec. \ref{sec:pdf}.

\begin{figure}[ht]
\centering
\includegraphics[width=0.98\linewidth,keepaspectratio]{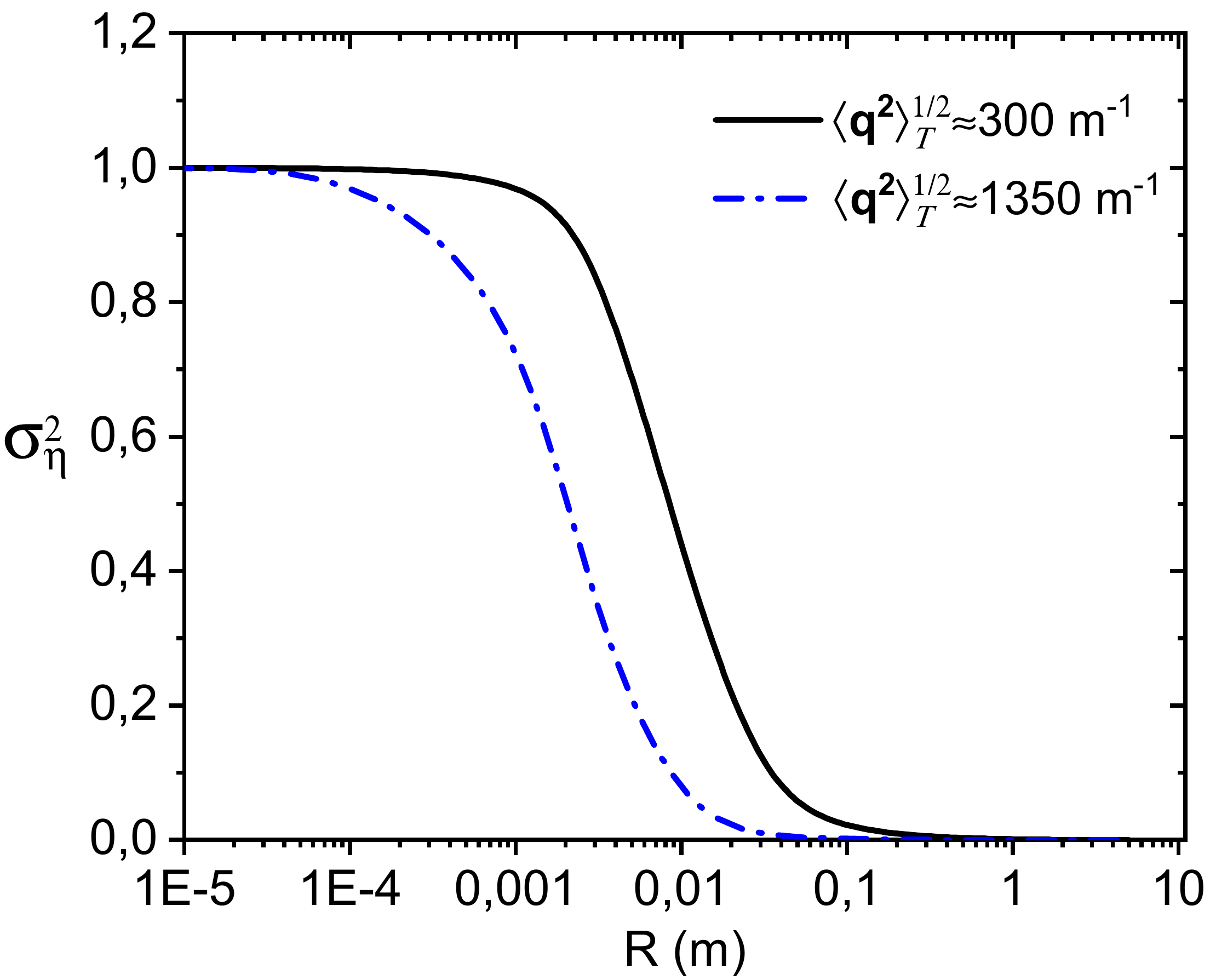}

\caption{ Aperture-averaged scintillation index vs. radius of receiver. Two curves represent propagation in two channels, gaining different change of transverse momenta of photons due to turbulence. Dash-doted line: $z=20\,km$, $C_n^2=2.5\times10^{-14}\,m^{-2/3}$; solid line: $z=100\,km$, $C_n^2=2.5\times10^{-16}\,m^{-2/3}$. Common parameters of the beam and channel for both curves: $r_0=0.01\,m$, $l_0/2\pi=10^{-3}\,m$, $q_0=10^{7}\,m^{-1}$. }
\label{fig:second}
\end{figure}

Numerical calculations illustrate the dependence of aperture-averaged scintillations on the radius of the detector aperture (see Fig. \ref{fig:second}).
There are two important features of fluctuation transmittance. The first feature relates to the  point-like aperture, in which the scintillation index tends to be unity, indicating the effect of saturation of fluctuations (see Sec. II). The second feature is the absence of 
fluctuations for apertures whose dimensions are much larger than the beam radius.

It can be seen in Fig. 3 that the fluctuation reduction in the asymptotic case strongly depends on the value $\langle{\bf q}^2\rangle$. A similar dependence on $R$ was observed  in earlier works. This phenomenon was also treated in terms of a coherence length for covariance function  \cite{Fante3, Churnside} that is similar to our explanations. 


The attempts to study fluctuations of light intensity in the range of of strong turbulence were undertaken  by Churnside \cite{Churnside} and Andrews \textit{et al} \cite{Andrews2}. To our knowledge, the rigorous derivation of analytical expression for $\sigma_{\eta}^2$ is given only in the present article.

\section{Discussion}
\label{sec:discussion}
The coarse-grained photon distribution function, being a specific type of the second moment of the light field, can adequately describe laser beams evolution. It follows from the definition of PDF (\ref{1threee}) that the wave-vector ${\bf k}$ describes the spatial distribution of the radiation field. Another wave-vector, ${\bf q}$, describes rather the kinetics of the PDF. There are opposite trends in the evolution of characteristic values of ${\bf k}$ and ${\bf q}$. 
Spatial and temporal changes correspond to an increase in the radius of the beam (resulting in the decrease of characteristic values of  ${\bf k}$)  and an increase in the speed of this process over time.

The solution of the averaged  kinetic equation for the case of the paraxial beam illustrates the significant anisotropy of the distribution function. The anisotropy is  due to photon drift with the velocity $c{\bf q}/q_0$ in the direction parallel to the vector ${\bf r}$. It follows from Eq. (\ref{45my}) that the center of spatial distribution of the beam moves with the velocity $c{\bf q}/(2q_0)$. A simple analytical  expression for PDF (\ref{45my}), obtained for long-distance propagation, confirms  our interpretation of the results shown in Fig. \ref{fig:zero}. 

The fourth moment  describes the intensity of light fluctuations. The corresponding noise level impairs the possibility of using laser beams for practical purposes. At the same time the fourth moment describes the non-local nature of photon fluctuations. This specific property of light is suitable for detailed diagnostics (scanning) of the beam body. Two detectors can be used for this purpose (see Sec II).

By solving the Langevin equation for fluctuating part of the PDF it is possible to obtain the fourth moment. To do this, it is sufficient to find the average of products of the obtained solutions. This program was implemented in our previous paper where a scintillation index was calculated. Unfortunately, the results relate only to weak and moderate turbulence due  to the use of iterative procedures there. In the general case, obtaining the fourth moment requires many integrations, which complicates the real calculations. At the same time, the problem simplifies in the case of light propagation over long distances where there is a regime of strong turbulence.  Randomization of the transverse momentum  coursed by light-eddies collisions changes the statistical properties to those where only pairwise correlations remain.

Using the above reasoning,  the  fourth moment for fluctuations  $\langle\delta\hat{I}({\bf r}, t)\delta\hat{I}({\bf r^\prime}, t)\rangle $ was expressed as the sum of linear and quadratic values of the mean PDF. Linear terms (the shot noise) describe the remote parts of the beam cross-section while the quadratic terms describe the usual (classical) fluctuations of radiations in the central area. The shot noise is the realization of quantum fluctuations. It reveals the discreet nature of light with low intensity (see Sec. V and the article by \cite{Kolobov}). It can be interpreted as a noise of non-interacting and non-correlated particles (quasi-particles). The distance where the classical noise  becomes equal to quantum noise is obtained analytically in the previous section. It depends on time $t$ and the  intensity of radiation.

Explicit fourth moment is used here for  obtaining the aperture averaging of fluctuations. The limited aperture of the receiver  is typical for most applications, so a rigorous analytical result for transmittance of light fluctuations can be useful. In the asymptotic case, the gain in transverse momentum  defines the correlation length for the fourth moment  The explicit value of the correlation length can be used for the optimal choice of the detector diameter.

The version of the kinetic equation with the "collision term" is  applicable to almost any propagation distances and strengths of the structure factor $C_n^2$. An alternative kinetic equation, in which the influence of a turbulence is considered in the framework of the ``random force" approach, is analyzed in the Appendix B. It is shown that for long-distance propagation, both approaches give identical results. This is due to relative decrease of the photon momentum transfer in individual photon-eddy collisions (``smooth collisions").   

\section{Conclusion}
The equation for the average PDF is solved analytically using the Fourier transform in spatial and momentum domains. It is shown that the asymptotic value of the fourth momentum of light field is expressed in terms of linear and quadratic in PDF forms  which can be easily calculated. The fourth moment describes the noise level of optical signals and can be used to estimate the performance of optical systems. In particular, they can be used to diagnose optical radiation similar to noise measurements used in the  fields of solid state physics and  surface science.  The results, obtained here, are almost independent on the configuration of the input radiation and could be applied to other optical systems.

\section{Acknowledgments}
The authors thank D. Vasylyev, A. Semenov, and E. Stolyarov for useful discussions and comments. The work of R.B. was partially supported by
grant(\#0120U100155) for Young Scientists Research Laboratories of the National Academy of Sciences of Ukraine.

\appendix

\section{Time dependence in paraxial approximation}
\label{appendix1}

When the paraxial beam propagates in the atmosphere, it is possible to replace in  the kinetic equation (18) the quantity $ct$ with $z$. To prove this, we multiply both sides of Eq. (\ref{18my}) by $c$, integrate over all values of  ${\bf r}_\bot$, and sum up over {\bf q}. As a result, we obtain for $I_T(z,t)$, which were introduced in Sec. \ref{sec:pdf}, the equation:
\begin{equation}\label{A0_1}
(\partial_t+c\partial_z)I_T(z,t)=0.
\end{equation}
There are no dissipative mechanisms in this equation. The total photon flux in the $z$-direction conserves regardless of the atmospheric turbulence. Any function  $I_T(z-ct)$ satisfies Eq. (\ref{A0_1}). At the same time,  this function must also satisfy the boundary and initial conditions given at $z=0$ and $t=0$. In this particular case equality $I_T(z-ct)=I_T(0)$ holds and one may   conclude that the function $I_T(z-ct)$ obeys the initial conditions only if $z=ct$.  This property of radiation applies only to cases of paraxial propagation. Our study concerns just this case. 

\section{Average PDF for Boltzmann collisionless equation}
\label{appendix2}
In this Appendix, we consider the applicability  of the collisionless Boltzmann kinetic equation for the description of beam propagation. The atmospheric turbulence in this equation is taken into account by substituting expression $-{\bf F}\cdot{\partial_{\bf r}}\hat{f}({\bf r},{\bf q},t)$ instead of $\hat{K}({\bf r},{\bf q},t)-\hat{\nu}_{\bf q} { \hat{f}({\bf r},{\bf q},t) } $ into the right-hand side of (\ref{13my}), we obtain an alternative kinetic equation in which turbulence is  manifested by a random force $ {\bf F}=\omega_0\partial_{\bf r}n({\bf r}) $ acting on photons. Such an equation is applicable for description of the range of moderate and strong turbulence (see, for example, Refs. \cite{berm,berm2009,baskov2016}).
\begin{equation}\label{A1} 
\{ \partial_t +{\bf c_q}\partial_{\bf r}+{\bf F}({\bf r})\partial_{\bf q}\}
\hat{f}({\bf r},{\bf q},t)=0.
\end{equation}
Our aim is to solve this equation and compare it with the solution given by Eq. (\ref{29my}). 
The general solution of (\ref{A1}) can be written as
\begin{equation}\label{A2}
\hat{f}({\bf r},{\bf q},t)=\phi\bigg({\bf r}-\int _0^tdt^\prime{\bf {\dot r}}(t^\prime);
{\bf q}-\int_0^tdt^\prime{\bf {\dot q}}(t^\prime)\bigg),
\end{equation}
where $\phi ({\bf r},{\bf q})$ is the photon distribution function in the aperture plane of the source: $f({\bf r},{\bf q},t=0)=\phi ({\bf r},{\bf q})$. In this case, $\phi $ is expressed via the operators $b^\dag_{{\bf q}+ \frac{\bf k}2}, b_{{\bf q}-\frac{\bf k}2}$  at $t=0$:  
\begin{equation}\label{A3}
\phi ({\bf r},{\bf q})=\frac 1V\sum_{\bf k}e^{-i{\bf kr}}(b^\dag_{{\bf
q}+ \frac{\bf k}2}b_{{\bf q}-\frac{\bf k}2})|_{t=0}\equiv \sum_{\bf
k}e^{-i{\bf kr}} \phi ({\bf k},{\bf q}).
\end{equation} 
Partial integration in Eq. (\ref{A2}) reduces it to 
\begin{equation}\label{A4}
\hat{f}({\bf r},{\bf q},t)=\phi \Big\{{\bf r}-{\bf c_q}t+\frac c{q_0}\int_0^t dt^\prime t^\prime {\bf F} [{\bf r}(t^\prime )];
{\bf q}-
\int _0^t dt^\prime {\bf F} [{\bf r}(t^\prime )]\Big\},
\end{equation}
where ${\bf c_q}$ is the photon velocity at time $t$. It is assumed that the propagation distance $z$  is equal to $ct$ and  the fluctuating force ${\bf F}$ as well as trajectories ${\bf r}({\bf q},t^\prime )$ are perpendicular to the $z$ axis. In this case, the problem reduces to considering the evolution of photons only in the $x,y$ plane.  

The explicit form of $\phi ({\bf k},{\bf q})$ can be obtained from matching conditions of the source field and the field in the atmosphere. The quantum amplitude $b_{{\bf q}}$  enter to the electromagnetic field via the product  $b_{{\bf q}}e^{i{\bf q}\cdot{\bf r}}$. It is assumed that the pattern of source field  in the aperture plane is given by function $\Phi({\bf r_\bot})$. Therefore, one may write  
\begin{equation}\label{A5}
b_{{\bf q_\bot},q_0}\propto C^\prime\int d{\bf r_\bot}e^{-i{\bf k_\bot r_\bot}}\Phi({\bf r_\bot}).
\end{equation}
Using Eq. (\ref{A3}) and considering 
$\Phi({\bf r_\bot})$ as a Gaussian function $\Phi({\bf r_\bot})= const\times e^{-{r^2_\bot}/r_0^2}$, we obtain 
\begin{equation}\label{A6}
\langle\phi ({\bf k},{\bf q})\rangle_{qm}=C^{\prime\prime}e^{-(q_\bot^2+k_\bot^2/4)r_0^2/2},
\end{equation}
where the symbol  $\langle...\rangle_{qm}$ indicates a quantum-mechanical averaging of operators in the angle brackets.
Coefficients $C^{\prime}, C^{\prime\prime}$ can be obtained if the total photon flux is known.   

Using Eqs. (\ref{A3}),(\ref{A4}) and (\ref{A6}), we find 
\begin{equation}\label{A7}
\langle \hat{f}({\bf r_\bot},{\bf q_\bot},t)\rangle_{qm}=C^{\prime\prime}\sum_{{\bf k}_\bot} e^{-i{\bf k_\bot R}(t)-Q^2(t)r^2_0/2-k^2_\bot r^2_0/8},
\end{equation}
where 
\[{\bf R}(t)={\bf r}_\bot-{\bf c_q}_\bot t+\frac c{q_0}\int
_0^tdt^\prime t^\prime {\bf F} [{\bf r_\bot}({\bf q_\bot},t^\prime )]\]
and
\[{\bf Q}(t)={\bf q_\bot}-\int _0^t dt^\prime {\bf F} [{\bf r_\bot}({\bf q_\bot},t^\prime )].\]
Further consideration is simplified if we use the identity
\begin{equation}\label{A8}
e^{-Q^2r_0^2/2}=\int \frac {d{\bf p}}{2\pi r_0^2}e^{-i{\bf
pQ}-p^2/(2r_0^2)},
\end{equation} 
where the integration is in the plane $S$. Using Eq. (\ref{A8}) we can exclude $Q^2$ in the exponent of the right side of (\ref{A7}). Than we obtain much simpler expression for the distribution in which only one of the exponents contains a linear in {\bf F} term: 
\begin{eqnarray}\label{A9}
\langle\hat{f}({\bf r}_\bot,{\bf q}_\bot,t)\rangle_{qm}{=}\frac {C^{\prime\prime}}{2\pi r_0^2}\sum_{\bf k}\int d{\bf p}e^{-i{\bf k_\bot}\{ {\bf r}{-}{\bf c_q}t\} -i{\bf p}{\bf q}}\nonumber\\
\times e^{-k^2r_0^2/8-p^2/2r_0^2}  e^{i\int _0^tdt^\prime\left({\bf p}-{\bf k}\frac{ct^\prime}{q_ 0}\right) {\bf F}[{\bf r}(t^{\prime})]}.
\end{eqnarray}

The last factor in (\ref{A9}) is simplified after averaging over the inhomogeneity of the atmosphere. Assuming the random quantity 
\[\Pi={i\int _0^tdt^\prime\left({\bf p}-{\bf k}\frac{ct^\prime}{q_ 0}\right) {\bf F}[{\bf r}(t^{\prime})]}\] 
obeys the Gaussian statistics,  we can write             
\begin{equation}\label{A10}
\langle e^\Pi\rangle=e^{\langle \Pi^2\rangle /2} 
\end{equation} 
where 
\begin{equation}\label{A11}
\langle \Pi^2\rangle=-\frac{8\pi\omega_{0}^{2}}{c}\int_0^t dt^\prime\int d{\bf k'_{\bot}}\psi({\bf k'_{\bot}})\bigg[\bigg({\bf p}-{\bf k_\bot}\frac{ct^\prime}{q_0}\bigg){\cdot} 
\frac{\bf k'_{\bot}}2\bigg]^2.
\end{equation}
The following relations were used for obtaining Eq. (\ref{A11}):  
\[\langle n({\bf r}_{t^\prime})n({\bf r}_{t^{\prime\prime}})\rangle=\int d{\bf k^\prime}e^{i{\bf k}^\prime_\bot({\bf r}_{t^\prime} -{\bf r}_{t^{\prime\prime}}) -ik^\prime_z(z^\prime-z^{\prime\prime})}\psi_{\bf k^\prime}\]
\begin{equation}\label{A12}
\approx\frac{2\pi}c\delta(t^\prime-t^{\prime\prime})\int d{\bf k^\prime}_\bot{\psi_{\bf k^\prime_\bot}},
\end{equation}
where $z^\prime=ct^\prime$,$z^{\prime\prime}=ct^{\prime\prime}$. The use of a delta-function in Eq. (\ref{A12}) becomes possible due to a great difference of photon velocities in the parallel and perpendicular  to the propagation path directions:
$c\gg cq_\bot/q_0.$

For  long-distance propagation, the sine function included in expression (\ref{26my}) can be replaced by its argument.In this case we have the equivalence of Eqs. (\ref{A11}) and (\ref{26my}): \begin{equation}
\frac12\langle \Pi^2\rangle\approx-\gamma({\bf k_\bot},{\bf p},t).
\end{equation}
Also, the average distribution functions obtained within alternative approaches becomes identical. Therefore the explicit value of the fourth moment (\ref{7my}), can be derived using either of the two.

\end{document}